**RESEARCH**  **Open Access**

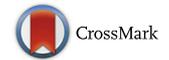

# Distributed and adaptive location identification system for mobile devices

Fahed Awad[1*], Aisha Al-Sadi[2], Fida'a Al-Quran[1] and Abdulsalam Alsmady[3]

**Abstract**

Indoor location identification and navigation need to be as simple, seamless, and ubiquitous as its outdoor GPS-based counterpart is. It would be of great convenience to the mobile user to be able to continue navigating seamlessly as he or she moves from a GPS-clear outdoor environment into an indoor environment or a GPS-obstructed outdoor environment such as a tunnel or forest. Existing infrastructure-based indoor localization systems lack such capability, on top of potentially facing several critical technical challenges such as increased cost of installation, centralization, lack of reliability, poor localization accuracy, poor adaptation to the dynamics of the surrounding environment, latency, system-level and computational complexities, repetitive labor-intensive parameter tuning, and user privacy. To this end, this paper presents a novel mechanism with the potential to overcome most (if not all) of the abovementioned challenges. The proposed mechanism is simple, distributed, adaptive, collaborative, and cost-effective. Based on the proposed algorithm, a mobile blind device can potentially utilize, as GPS-like reference nodes, either in-range location-aware compatible mobile devices or preinstalled low-cost infrastructure-less location-aware beacon nodes. The proposed approach is model-based and calibration-free that uses the received signal strength to periodically and collaboratively measure and update the radio frequency characteristics of the operating environment to estimate the distances to the reference nodes. Trilateration is then used by the blind device to identify its own location, similar to that used in the GPS-based system. Simulation and empirical testing ascertained that the proposed approach can potentially be the core of future indoor and GPS-obstructed environments.

## 1 Introduction

The need for GPS-like ad hoc indoor navigation and location identification systems is a tenacious fact nowadays. It is very annoying to lose the capability to navigate or identify your current location as soon as a building or a tree shadows the GPS signal. Imagine having a mobile device that is able to continue navigating seamlessly as you enter a large public building such as a shopping mall, hospital, or government complex without any user intervention. Alternatively, imagine being shadowed from the GPS signal, but still being able to identify your location seamlessly using other mobile devices as reference points. Similarly, during a search-and-rescue operation, the mobile device or a trapped victim can identify its own location using the rescuers' mobile devices as reference points and send help beacons with its coordinates [1].

As mobile devices are becoming an integral part of people's daily lives for the countless number of applications and services they provide, more and more mobile-based location-based services and solutions are yet to be discovered and developed. In fact, mobile devices are no longer used only for conventional communication services, but also to provide advanced sensing capabilities due to powerful sensors equipped within, including motion sensors (e.g., gyroscope and accelerometer) and location sensors (e.g., GPS and Wi-Fi). Hence, the valuable sensed location can be leveraged in many different location-dependent applications and research projects such as mobile-based travel surveys for urban planning [2, 3], tracking systems [4], and environmental solutions to estimate emission of $CO_2$ [5, 6]. Indeed, there are endless potential solutions for existing real-life problems.

* Correspondence: fhawad@just.edu.jo
[1]Department of Network Engineering and Security, Jordan University of Science and Technology, P. O. Box 3030, Irbid 22110, Jordan
Full list of author information is available at the end of the article

Recently, research communities have been very motivated to provide indoor mobile-based localization systems [7–10]. Enormous number of studies leverage location-based services in a wide range of applications with highly in-demand features. For example, location-based services are integrated with the Internet-of-Things (IoT) to track patients and staff in health care facilities [11] and track blind and impaired people around [12]. Additionally, most popular applications of location-based services are for indoor navigation purposes in large buildings such as museums and shopping malls.

For indoor localizations, the most popular and widespread technology used is Wi-Fi because it has become a commodity in every mobile device made nowadays. This reduces the cost and allows for leveraging the existing Wi-Fi infrastructure devices (i.e., base stations or access points) or other peer devices (i.e., other client devices in ad hoc mode) as reference nodes to obtain position information. Thus, using Wi-Fi for indoor localization represents a great opportunity to investigate efficient and low-cost solutions. Even though Wi-Fi may achieve acceptable performance, yet it still poses many limitations and challenges.

This work presents a novel approach to address one of the most challenging scenarios in wireless location identification, which is when the device that needs to identify its own location (called blind node) is shadowed from the GPS signal and no Wi-Fi infrastructure signal is in reach. In this case, it is sufficient to have at least three other in-range devices with known locations (called reference nodes) broadcast special beacon packets to help the blind node identify its own location. This beacon packet must contain the necessary information the blind node needs in order to estimate its own location in a seamless and dynamic way regardless of the surrounding environment. The reference nodes need not be interconnected in any specific way and can be other mobile devices, of which the GPS signal is not shadowed (e.g., an indoor mobile blind node using outdoor mobile reference nodes), or intentionally installed beacon nodes at known location (e.g., low-cost Wi-Fi IoT modules).

For example and without loss of generality, in emergency situations, where the communication infrastructure is either dismantled by catastrophes or does not originally exist (e.g., as in forests), search-and-rescue operations can make a crucial benefit from victims' smart mobile devices working in an ad hoc mode. When a victim is trapped under rubble or lost in the pushes, where no GPS signal is reachable, the rescuers' and the victim's mobile devices can collaborate an ad hoc Wi-Fi network mode to identify the victim's location [1]. In such case, the rescuers' mobile devices can act as reference nodes, forming GPS-like localization system. If the rescuers' devices happen to be in clear areas, where GPS signals are detected, an absolute identification of the victim's location is straightforward. Otherwise, a relative location identification is a good alternative, where the reference nodes are positioned at known locations and the location of the blind node is identified relative to these locations. Note that drones equipped with GPS receivers and Wi-Fi radios can also be used as reference nodes by flying above obstacles and providing reference to blind nodes.

In fact, the widespread usage of mobile devices led to the concept of smart environments [13], which means that more and more of wireless technologies and sensing capabilities are integrated with the environment to make it smart. Exploiting existing mobile devices in the environment to collaboratively assist in location identification systems is an attractive solution that could expand the spectrum of location identification services and seamlessly obtain location information from almost any environment even if it is not equipped with infrastructure devices.

Many indoor localization algorithms were proposed in the literature to enable location-based services. The general classification of existing mechanisms is calibration-free and calibration-based (also called model-based and fingerprinting-based, respectively). These mechanisms vary in accuracy, cost, and complexity, as well as robustness to the environment changes. In calibration-based methods, collected data set in the target environment are used to infer an accurate relationship between each position and the corresponding received signal strength (RSS). This type requires prior knowledge of the target environment, and it is labor-intensive. Relatively high levels of accuracy can be achieved in calibration-based localization, but it required periodic recalibration, especially in dynamic environments. On the other hand, calibration-free are less complex and can be smoothly configured based on the characteristics of the environment, which makes it more practical for ad hoc mode. However, this requires an accurate and dynamic estimation of the environment characteristics.

Wi-Fi-dependent location identification techniques are generally based on measurements of time-of-arrival (TOA) [14], time-difference-of-arrival (TDOA) [15], angle-of-arrival (AOA) [16–18], or RSS [19]. Both TOA and TDOA methods suffer from time synchronization problems. Therefore, most existing Wi-Fi location identification solutions are based on AOA or RSS. Although, AOA-based methods can achieve higher accuracy than RSS-based, it is more complex and it requires a special type of antennas [20]. The major advantages of RSS-based methods are its low complexity and speed of calculation compared to the other methods [14–16, 18]. Thus, RSS-based methods are the most widely used [19–22].

In general, RSS-based methods are highly dependent on the environmental characteristics and changes, which can degrade the overall accuracy. Therefore, the accuracy of RSSI-based methods is still not satisfactory due to the frequent changes of RSS level even in non-changing environments [23]. As a result, more development needs to be investigated to further enhance the level of accuracy.

Many researchers have attempted to validate and improve RSS methods from different perspectives. In [24–26], fusion of additional information obtained from body-mounted sensors of the mobile devices asset the accuracy level. However, these approaches suffer from high computational power and thus degrade the performance of the localization system. In [27], an AI algorithm was proposed to adjust the location of existing access points until each specific area receives a unique set of RSS values and thus enhance the accuracy of RSS-based localization algorithms. The algorithm was shown to achieve 90% of RSS uniqueness when seven access points are active. Hence, this strategy is suitable for large-scale environments with a large number of access points.

Calibration-based solutions use fingerprints of average RSS values over a specific duration at specific locations in order to diminish the impact of variation in the RSS values [28–30]. Even though calibration-based techniques provide more localization accuracy than calibration-free, they require extensive efforts to collect the calibration data. This problem rises dramatically in dynamic environments since any change in the environment or the configuration, requires new calibration data to be collected. To address this issue, a number of methodologies were proposed to reduce the time required to collect the calibration data. For example, in [31], an effortless indoor Wi-Fi solution that leverages the floor plan and walls within the environments to generate RSS maps of access points and the target node location is estimated using a map overlapping technique. This work eliminates the necessary time to survey the environment. However, it does not address the changes of RSS due to changes in the environment over time, to provide accurate results. In [32], a novel algorithm that solves the problem of regenerating calibration dataset was proposed. In this approach, access points with custom-made firmware were allowed to scan the channel and record RSS levels of each other, as well as the RSS level of a special anchor device. It also requires the mobile device to be in an access point mode in order to send beacon frames and allow other nodes in the system to connect to it. Although, in [31, 32], good attempts were made to solve the limitations of fingerprinting methods, these are client-server-based techniques and cannot be adapted for distributed Wi-Fi location identification systems.

Calibration-free solutions are usually based on signal propagation models, thus called model-based techniques. The distinction among these techniques lays in the way the parameters of the used model(s) are inferred in order to estimate the distance to the reference nodes. In [21], a blind node sends the RSS levels obtained from the beacon frames of the existing access points to a remote server. This server dynamically adjusts the propagation model based on the RSS values and localizes the blind node using trilateration. The reported mean localization error is slightly lower than 4 m. "A self-adaptive model-based Wi-Fi indoor localization method" introduced in [33] used extended versions of two well-known propagation models (namely the free-space path loss and ITU). Similar to [32], access points with custom-made firmware were used in monitor mode to capture each other's beacon frames. The collected data are transferred to a centralized server that continuously estimates the propagation parameters of the environment and the location of the mobile terminal. The method requires the positions of access points and positions and properties of the walls. The localization accuracy was in the range of 2 to 4 m. The "multiple frequency adaptive model-based indoor localization method" (MFAM) proposed in [34] was built upon the work in [33] using multiple frequency bands. MFAM was reported to improve the localization performance of its predecessor by 6%. The authors of [35] utilized a two-slope channel model to propose a robust indoor mobile target tracking algorithm using a set of interacting multiple models, each involving two extended Kalman filters. The target mobile node collects the RSS levels and sends them to a centralized tracking server. The system was tested using a mobile robot, and the reported mean localization error was 0.19 m. Thus, all abovementioned RSS-based calibration-free methods are, in fact, localization (i.e., tracking) rather than location identification techniques that depend primarily on infrastructure devices and a central server to estimate the parameters of the model and calculate the estimated location. To address this issue, we proposed a novel approach called "smartphone-assisted location identification (SALI)" for search-and-rescue services [1], which was the first collaborative calibration-free RSS-based ad hoc Wi-Fi location identification system.

In this paper, we present an RSS-based, distributed, calibration-free, and real-time location identification strategy for mobile devices, called distributed and adaptive location identification system (DALIS), as a comprehensive extension of SALI algorithm. To this aim, we performed a comprehensive study on the propagation model parameters that influence the accuracy of RSS-based location estimation. Based on that, a distributed algorithm that adapts to the dynamics of the changing environment characteristics was devised. To the best of our knowledge, this is the first attempt to investigate the possibility of using RSS-based location identification in a distributed ad hoc fashion even when reference nodes are mobile. According to the simulation and experimental testing results, the proposed mechanism is robust enough with both mobile and stationary nodes. As a result of this study, we propose an efficient low-cost and infrastructure-less indoor location identification strategy using IoT Wi-Fi modules.

## 2 Methods

DALIS is based on the lognormal shadowing with exponential path loss model for wireless signal propagation [36, 37], which fits both indoor and outdoor environments since it can be configured according to the corresponding environmental characteristics. This model is expressed as follows:

$$P_{RX} = P_{TX} - PL(d_0) - 10\, n\, \log_{10}(d/d_0) + \chi_\sigma \quad (1)$$

where:

- $P_{TX}$ is the transmitted power at the transmitting antenna in dBm.
- $PL(d_0)$ is the path loss at some reference distance $d_0$ from the transmitting antenna in dB. Usually, $d_0 = 1$ to 10 m for indoor environments.
- $n$ is the path loss exponent (PLE), which is dependent on the specific propagation environment.
- $\chi_\sigma \sim N(0, \sigma^2)$ is a normally distributed random variable with zero mean and standard deviation $\sigma$. $\chi_\sigma$ represents the variation in RSS caused by the random shadowing.

Note that $PL(d_0)$, $n$, and $\chi_\sigma$ are environment-dependent parameters that are experimentally measured. However, typically, $P_{TX}$ is between 15 and 20 dBm and, for indoor environments, $PL(d_0 = 1\text{ m})$ is around 40 dB and $\chi_\sigma$ converges to zero after averaging sufficient number of RSS samples at the same location with $\sigma$ between 1 and 8 dB [37]. On the other hand, $n$ is the most critical parameter since it represents the exponential decay of the signal as a function of the distance and it may dynamically vary around the clock according to the instantaneous changes in the environment. Therefore, $n$ needs to be updated on real time in order to obtain an accurate estimation of the signal and distance. Thus, given $d$, $P_{TX}$, and $PL(d_0)$, an instantaneous estimation of $n$, or $\hat{n}$, can be calculated. Similarly, given $P_{TX}$, $PL(d_0)$, and $\hat{n}$, an estimation of $d$, or $\hat{d}$, can be calculated as follows:

$$\hat{n} = \frac{P_{TX} - PL(d_0) - P_{RX}(d) + \chi_\sigma}{10\, \log_{10}(d/d_0)} \quad (2)$$

$$\hat{d} = d_0 \times 10^{\frac{P_{TX} - PL(d_0) - P_{RX}(d) + \chi_\sigma}{10\hat{n}}} \quad (3)$$

The main principle of the current approach is that there must be at least three reference nodes within communication range of each other and that each reference node must periodically broadcast a beacon packet that contains its current coordinates, $P_{TX}$, and its $\hat{n}$. Each reference node obtains the RSS values associated with the beacon frames from the reference nodes, along with their coordinates, and calculates the Euclidean distance, the average RSS, and $\hat{n}$. This allows each reference node to keep $n$ dynamically updated and announced on real time. On the other hand, for a blind node to be able to estimate its own location, it must receive beacon packets from at least three reference nodes. This allows the blind node to estimate the distance to each of the reference nodes and then use the trilateration method [38] to identify its own location.

The operation of DALIS is summarized as follows:

1. Reference node $RN_i$ periodically sends a beacon packet, which includes the following parameters:

- Its transmit power, $P_{TXi}$
- Its current coordinates, $(x_i, y_i)$
- Its current estimation on $n$, $\hat{n}_i$

2. After receiving $w_r$ beacon packets from $RN_i$, reference node $RN_j$ calculates the Euclidean distance with $RN_i$ as follows:

$$d_{ji} = \sqrt{(x_i - x_j)^2 + (y_i - y_j)^2} \quad (4)$$

and calculates the average RSS from $RN_i$ and the estimated $n$ from $RN_i$, $\hat{n}_{ji}$ using (2) as follows:

$$\tilde{P}_{RXi} = \frac{1}{w_r} \sum^{w_r} P_{RXi} \quad (5)$$

$$\hat{n}_{ji} = \frac{P_{TXi} - PL(d_0) - \tilde{P}_{RXi} + \frac{1}{w_r}\sum^{w_r}\chi_\sigma}{10\, \log_{10}(d_{ji}/d_0)} \quad (6)$$

where $\tilde{P}_{RXi}$ is the average of the last $w_r$ RSS levels obtained. Note that $w_r$ is the window size of the moving average of RSS. For sufficient $w_r \gg 1$,

$$\frac{1}{w_r}\sum^{w_r}\chi_\sigma = E[\chi_\sigma] = 0 \quad (7)$$

so,

$$\hat{n}_{ji} = \frac{P_{TXi} - PL(d_0) - \tilde{P}_{RXi}}{10\, \log_{10}(d_{ji}/d_0)} \quad (8)$$

3. Reference node $RN_j$ calculates the overall estimated $n$ from all reference nodes, $\hat{n}_j$, by averaging $\hat{n}_{ji}$ of all other reference nodes within communication range as follows:

$$\hat{n}_j = \frac{1}{N_j} \sum_{i=1}^{N_j} \hat{n}_{ji} \quad (9)$$

where $N_j$ is the number of reference nodes within communication range of $RN_j$.

4. After receiving $w_r$ beacon packets from $RN_i$, the blind node calculates the average RSS from $RN_i$ as in (5) and estimates the distance to $RN_i$ using (3) as follows:

$$\hat{d}_i = d_0 \times 10^{\frac{P_{TX_i} - PL(d_0) - \tilde{P}_{RX_i}}{10 \hat{n}_i}} \quad (10)$$

then calculates the average distance to $RN_i$ for the last $w_d$ estimations of $\hat{d}_i$ as follows:

$$\tilde{d}_i = \frac{1}{w_d} \sum^{w_d} \hat{d}_i \quad (11)$$

5. After calculating $\tilde{d}_i$ for three reference nodes, the blind node estimates its location, $\hat{L}$, using the trilateration method [38]. Then, take the average (i.e., the centroid) of the last $w_l$ location estimations as follows

$$\tilde{L} = \frac{1}{w_l} \sum^{w_l} \hat{L} \quad (12)$$

$\tilde{L}$ represents the current estimated location of the blind node.

The reader might wonder about the reason we define averaging windows rather than averaging the samples altogether. One main reason is that averaging all the obtained samples works only when all nodes, whether reference or blind, are stationary, but if any node is mobile, the sample becomes obsolete as soon as the nodes move away from the corresponding location. In fact, having such samples in the average calculation degrades the localization performance, as we will show in the results section.

## 3 Simulation results

The performance of DALIS was evaluated statistically via simulation using QualNet network simulator [39]. The performance metric used throughout this study is the mean localization error (MLE), which is the Euclidean distance between the blind node's actual location and the estimated location. First, we investigate the effect of averaging the RSS and the location estimation on the localization accuracy. Then, we study several possible practical scenarios, ranging from all nodes being stationary to all nodes being mobile. For selected scenarios, we study the effect of the roaming area, the averaging window sizes, and the standard deviation of the RSS signal fluctuation. Additionally, for the mobile scenarios, we study the impact of the speed of mobility.

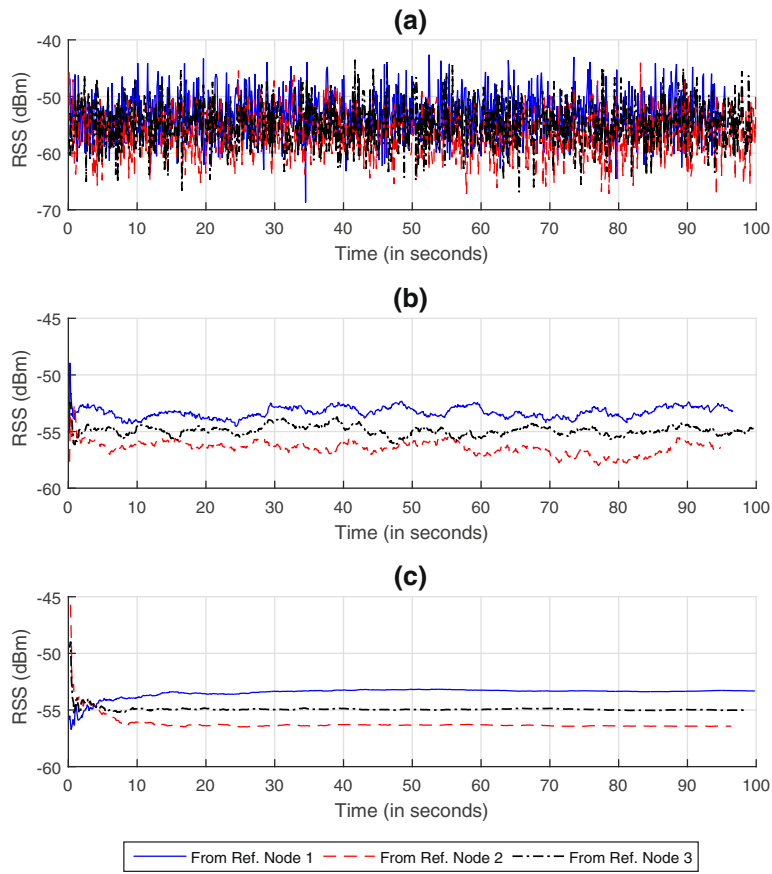

**Fig. 1** Effect of averaging RSS by the blind node when all nodes are stationary. (**a**) No averaging (i.e., only instantaneous RSS sample is considered), $w_r = 1$. (**b**) Intermediate window size, $w_r = 50$. (**c**) Average all RSS samples, $w_r = \infty$

### 3.1 Case 1: Stationary nodes

The case of stationary nodes may not be the most practical, but it helps us understand the effect of different parameters on the performance of the current mechanism. Since it represents the best case possible for localization efficiency, it allows us to define the upper performance limits.

We investigated the impact of $w_r$ and $w_l$ on the localization performance when all nodes are stationary by examining three window sizes for each: the minimum (i.e., no averaging), which is the worst case; an intermediate (i.e., averaging a fair number of samples), which is the average case; and the maximum (i.e., averaging all received samples o far), which the best case.

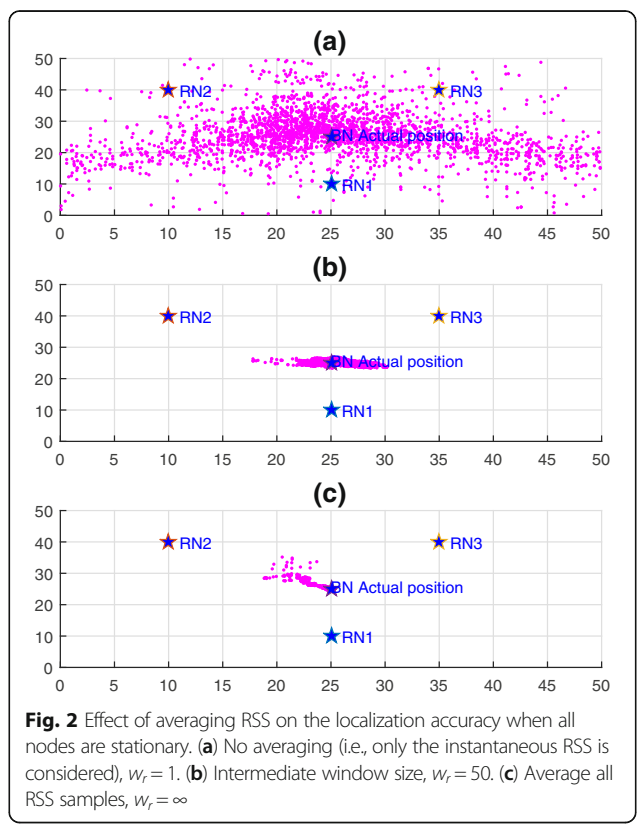

**Fig. 2** Effect of averaging RSS on the localization accuracy when all nodes are stationary. (**a**) No averaging (i.e., only the instantaneous RSS is considered), $w_r = 1$. (**b**) Intermediate window size, $w_r = 50$. (**c**) Average all RSS samples, $w_r = \infty$

**Table 1** A summary of the effect of average window size on the mean localization error. Each result is the average MLE of 100 trials, each with a different random seed

| Parameter | $w = 1$ | $w = 50$ | $w = \infty$ |
|---|---|---|---|
| RSS | 10.33 m | 1.71 m | 0.71 m |
| Distance | 10.33 m | 3.39 m | 1.83 m |
| Location | 10.33 m | 2.88 m | 0.94 m |

#### 3.1.1 Effect of averaging window sizes

Consider a 50 m × 50 m area with one blind node at the center and three reference nodes around the blind node. All nodes are stationary and operate based on DALIS. Each reference node broadcasts beacon packets at a rate of 10 packets per second with $\sigma = 4$ dB.

Figure 1 depicts the effect of RSS average window size on the stability of the signal level over time, as detected by the blind node for the three reference nodes. Figure 1a shows how much the instantaneous RSS level fluctuates with respect to time when no averaging is performed. It is obvious that the signal level is very unstable. Figure 1b shows that averaging the last 50 RSS levels obtained (i.e., over the past 5 s) improves the RSS stability quite a bit with some small-scale fluctuation left. In Fig. 1c, the advantage of averaging all RSS levels received so far is significant as the signal levels become nearly constant in a few seconds. Thus, the more RSS samples are averaged, the more stable the signal level becomes. However, this only applies when the sending and receiving nodes are stationary and the surrounding environment is mainly stable.

Figure 2 shows the effect of the RSS averaging and received signal on the localization accuracy and distribution of the estimated location around the actual location. Each dot represents one estimated location. Note that the first few estimated locations are always the worst since only a few RSS samples are used.

To investigate the individual effect of each parameter's average window size on MLE, we ran 100 experiments per scenario, each with a different random seed. Table 1 lists the average MLE of each scenario. It is obvious that averaging the RSS samples provides the best noise filtration, followed by averaging the estimated locations, and the least is averaging the estimated distances. The reason is that the impact of the noise on RSS is linear, whereas it is exponential on the distance. On the other hand, even though the location estimation is dependent on the distance estimation, the estimated location tends to scatter around the actual location, which makes the centroid of the estimated locations always within near proximity of the actual location.

**Table 2** A summary of the effect of averaging RSS, estimated distances, and estimated locations altogether on the mean localization error for different window sizes and different standard deviations. Each result is the average MLE of 100 trials, each with a different random seed

| $\sigma$ (dB) | $w = 1$ | $w = 10$ | $w = 50$ | $w = 100$ |
|---|---|---|---|---|
| 2 | 6.38 m | 1.51 m | 0.78 m | 0.63 m |
| 4 | 10.33 m | 3.11 m | 1.55 m | 1.28 m |
| 6 | 11.18 m | 4.85 m | 2.35 m | 1.97 m |

Table 3 The percent improvement of averaging RSS, estimated distances, and estimated locations altogether on the mean localization error compared to the case with no averaging based on Table 2

| σ (dB) | w = 10 | w = 50 | w = 100 |
|---|---|---|---|
| 2 | 76% | 88% | 90% |
| 4 | 70% | 85% | 88% |
| 6 | 57% | 79% | 82% |

Table 4 The incremental percent improvement of averaging the parameters on the mean localization error compared to the case with no averaging based on Table 2

| σ (dB) | w = 10 | w = 50 | w = 100 |
|---|---|---|---|
| 2 | 76% | 48% | 18% |
| 4 | 70% | 50% | 17% |
| 6 | 57% | 52% | 16% |

It is worth noting that the computational complexity of estimating the location by the blind node per iteration is relatively very low, which makes it very practical for walking-speed real-time applications such as in-building navigation or efficient and timely saving of a trapped victim.

### 3.1.2 Effect of combined averaging

The results in Table 1 demonstrate that averaging one of the parameters can improve the localization performance considerably, especially either RSS or the estimated locations. Hence, it became worth investigating the possibility of combining the averaging of all parameters to investigate the potentially leveraging the benefit of all. Therefore, we repeated the same experiment as in the previous section, but with all parameters averaged at different window sizes

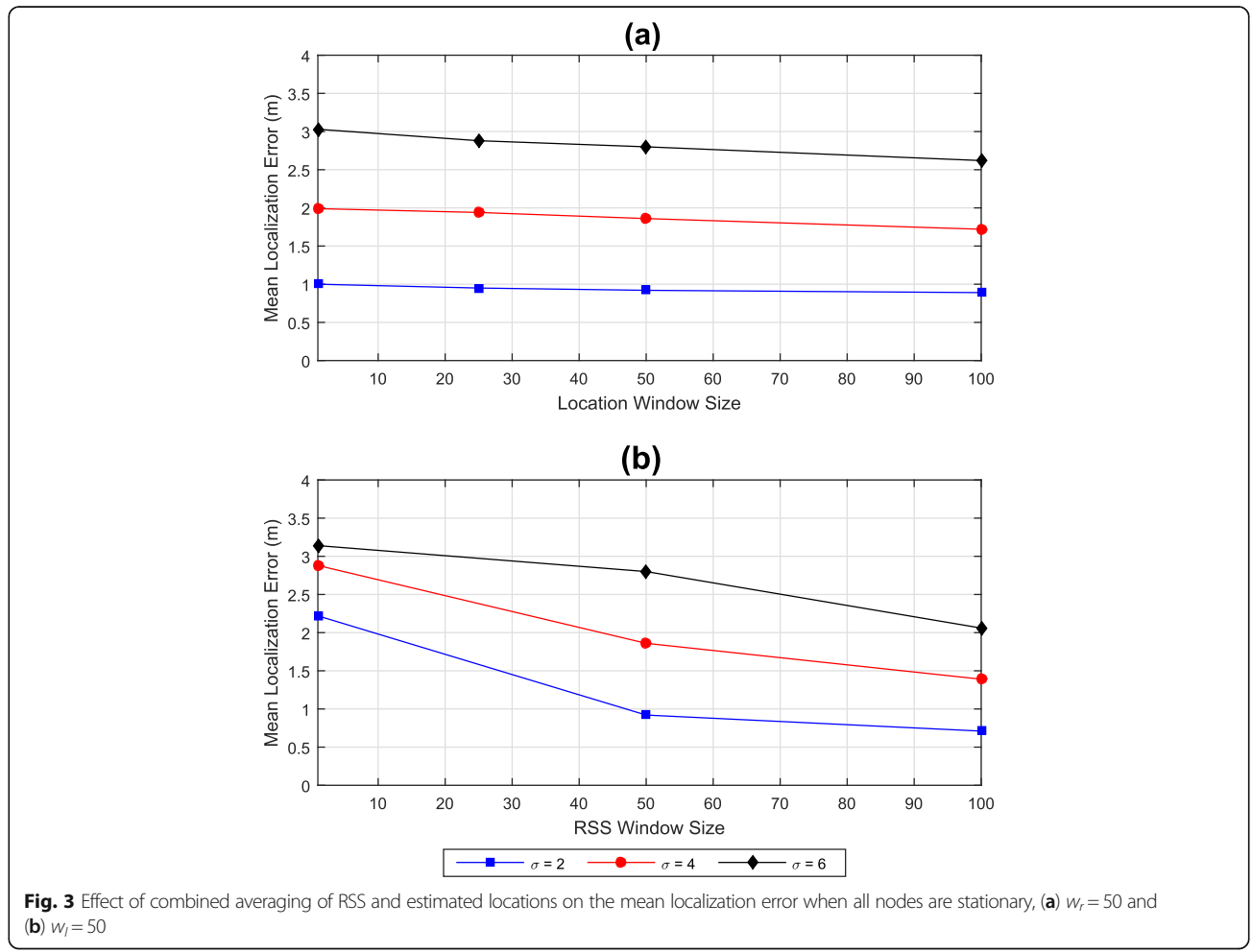

Fig. 3 Effect of combined averaging of RSS and estimated locations on the mean localization error when all nodes are stationary, (a) $w_r = 50$ and (b) $w_l = 50$

and standard deviations as listed in Table 2. The results in Table 2 suggest the following:

1. The potential to achieve decent localization accuracy is high, even with relatively small window sizes (i.e., $w \leq 50$). This indicates that a decent localization accuracy can potentially be achieved within a relatively short time, which is especially important when some nodes are mobile. That is, as the user is moving around at the walking speed, his/her mobile device can potentially track his/her movements within a couple of seconds. This is well investigated in the next section.

2. The percent improvement on the localization accuracy as a function of window sizes, compared to the case with no averaging, increases rapidly at the small window sizes, but it slows down noticeably as the window size increases, as listed in Table 3. That is, the incremental benefit of increasing the window size diminishes rapidly, as seen in Table 4. For example, increasing the window size from 10 to 50 improves the performance by 48%, whereas increasing it from 50 to 100 increases the performance only by 18%.

Now, since the averaging of the estimated distance is found to be much less effective than averaging the RSS and the estimated locations and for the purpose of reducing the computations on the blind node without impacting the localization performance, we investigated the possibility of skipping it. Thus, we performed two experiments, where in one experiment, we set $w_r = 50$ and changed $w_l$, whereas in the other experiment, we set $w_l = 50$ and changed $w_r$. Figure 3 shows the results of the two experiments, respectively. The most important observation in the figure is that when compared to Table 2, the contribution of averaging the estimated distance in improving the localization accuracy is less than 5%. This indicates that averaging the estimated distance can be an added cost with very small benefit.

### 3.1.3 Effect of area size and inter-node distance

We ran an experiment to investigate the relative impact of the area size and the corresponding separation among the nodes on the localization performance. We set $w_l = 50$, $\sigma = 4$ dB, and changed the area size and, correspondingly, the distances among the nodes. Figure 4 shows that the closer the reference nodes to the blind node, the better the localization accuracy. One reason is that as the farther the nodes get apart, the RSS becomes weaker and more prone to noise, interference, and multipath fading impairments, which corrupt the received signal level and increase the packet error rate (i.e., lower effective packet rate).

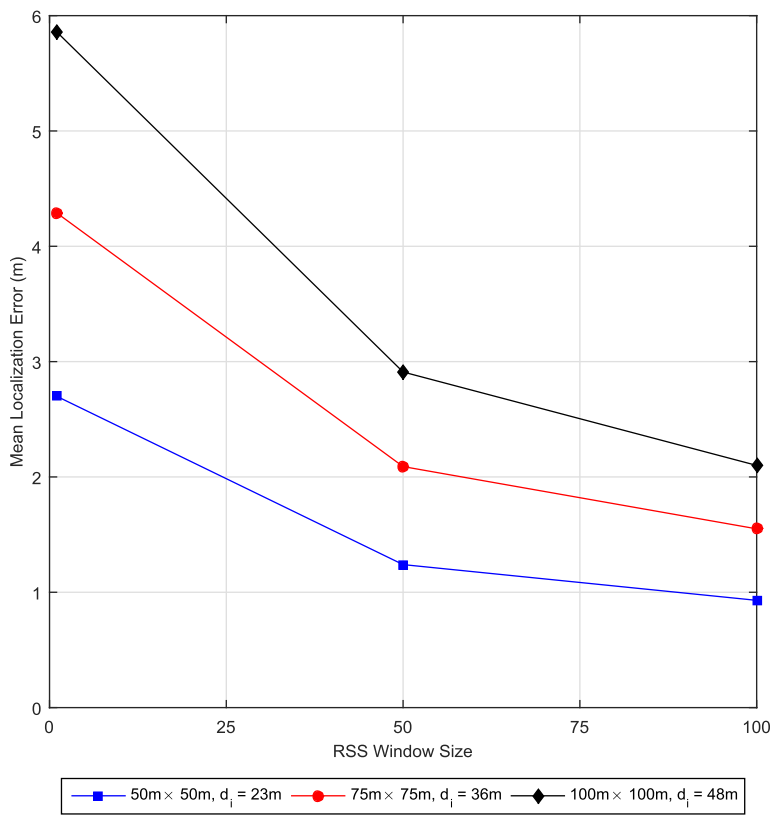

Fig. 4 Effect of area size and inter-node distance on the mean localization error when all nodes are stationary

## 3.2 Case 2: Mobile nodes

In this section, we investigate a more practical case, when some or all nodes are mobile. The objective of this part is to find out whether it is feasible to rely on mobile reference nodes. Reference node mobility limits the use of RSS averaging, and blind node mobility limits the use of location averaging. As the node moves, the distance $d_{ji}$ may no longer be constant in (4), (5), (6), and (8) and (7) may no longer converge to zero. However, at the walking speed of the person carrying the mobile device, the slight change in $d_{ji}$ within a short period in time may not have significant impact on the localization accuracy, given that the averaging window is small enough to minimize the impact and large enough to filter out the severe signal fluctuations, at least. The same thing applies to the location averaging when the blind node is mobile. It may not be very easy to optimize these parameters on real time, but we believe it is worth investigating the limits of such problems. To this end, we ran a number of experiments to test the limits of the proposed mechanism.

### 3.2.1 Mobile reference nodes and a stationary blind node

In this experiment, we investigate the effect of average window sizes on the localization accuracy when the reference nodes are mobile, which is the case of collaborative mobile ad hoc location identification. We first start with a stationary blind node, followed by the case when all nodes are mobile. The mobile nodes are assumed to follow the random way point (RWP) model [40] with a random average speed of 2 m/s (i.e., walking speed). In order to emulate the human walking behavior in an obstructed environment, where there are usually more turns than unobstructed environments, we modified the RWP model slightly by making the selected next location to move to within 10% of the area.

Figure 5 shows the results of allowing the reference nodes to move randomly at an average walking speed following RWP with zero pause time (i.e., continuous mobility), while the blind node is stationary at the center of a 50 m × 50 m area. Note that there is no location averaging performed by the blind node. It is obvious that the RSS samples can be beneficial for a short period of time before they expire or else degrade the performance if used after that. For this experiment, the optimal RSS average window size is nearly 30, which indicates that within 3 s of continuous random mobility (or 6 m of roaming around any starting point), averaging RSS levels is of benefit. Since the blind node is stationary, we repeated the same experiment, while the RSS average window size is set to 30 and averaging the estimated location. Figure 6 shows that this can improve the accuracy considerably over time.

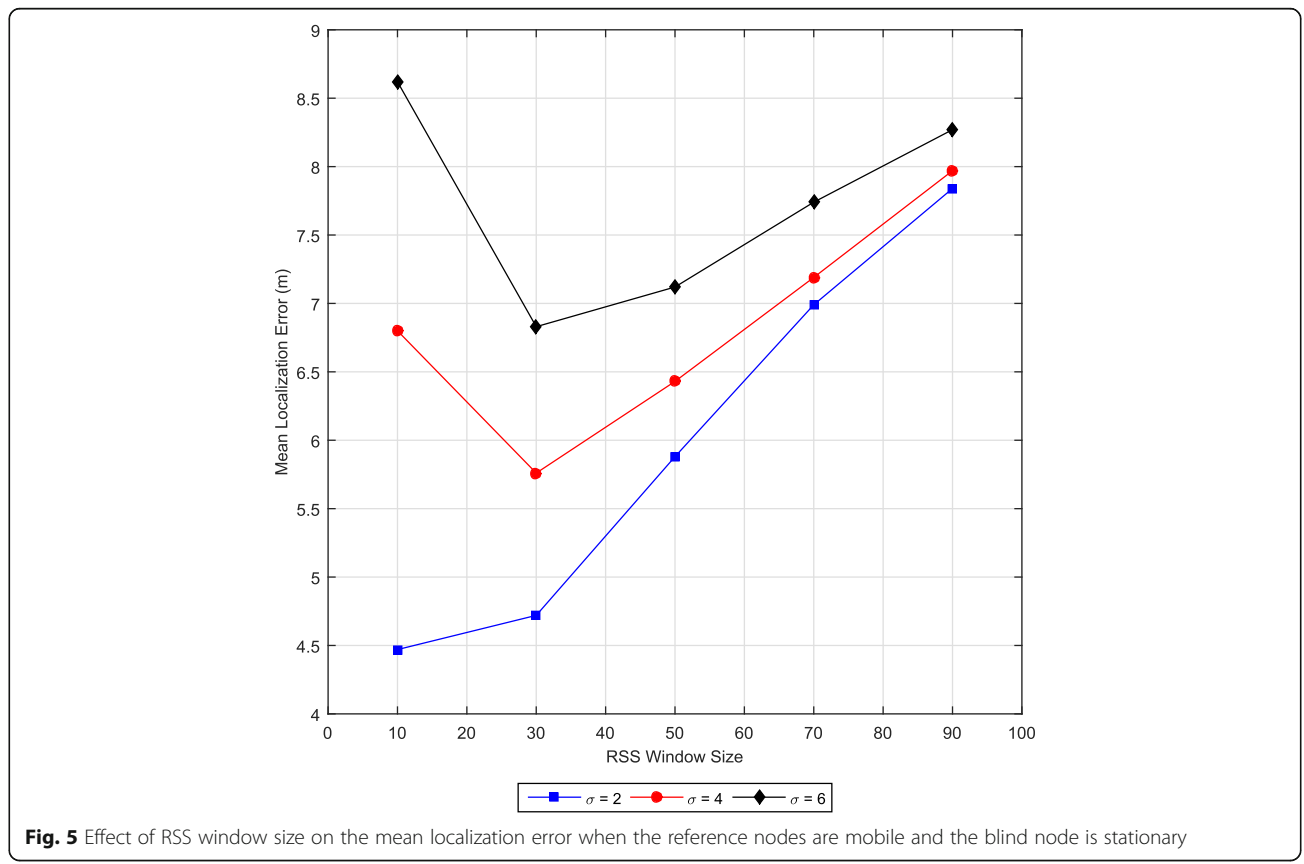

Fig. 5 Effect of RSS window size on the mean localization error when the reference nodes are mobile and the blind node is stationary

#### 3.2.2 Mobile reference nodes with a mobile blind node

In this experiment, we investigate the performance of the proposed mechanism in a large-scale collaborative mobile ad hoc mode. The mobile blind node attempts to identify its own GPS-shadowed location using mobile reference nodes roaming within a large area of 150 m × 150 m at an average random speed of 2 m/s and with RSS standard deviation of $\sigma = 4$ dB. All nodes are assumed to follow the RWP model, but with different pause time periods. Since the blind node is mobile, no location averaging is performed. The results are depicted in Fig. 7, which indicates that if the blind node pauses occasionally, a localization accuracy of 8 to 9 m can be achieved. This may not be good enough for regular indoor navigation, but it can be of a great benefit in search-and-rescue applications, especially a lost person in the bushes, where the trapped person may become within the visual or audible range. As seen earlier in Fig. 4, if the reference nodes (e.g., the rescuers in this case) move towards the blind node and become within closer proximity, the localization accuracy improves incrementally as reported in [1].

### 3.3 Case 3: Distributed low-cost and efficient indoor location identification system

Based on the observations we had on the results presented in the previous sections and given the recent IoT advancements and the associated low-cost and small form-factor System-On-Chip (SoC) devices with embedded Wi-Fi radios (e.g., ESP8266 module [41]), we propose a distributed low-cost and efficient indoor location identification system. These Wi-Fi SoC modules can be programmed to operate as beacon nodes based on DALIS strategy and be installed on the ceiling and/or sidewalls of large-scale indoor environments in a proper density to provide GPS-like referencing to mobile devices. That is, each module is set with hardcoded in-building coordinates, at which it is installed, and is programmed to exchange beacon packets with the other modules within its communication range. This allows all modules to experimentally and dynamically extract the RF environmental characteristics on real time and provide location referencing with reasonable accuracy. To prove the concept, we ran an experiment with a grid of 100 beacon nodes using different grid sizes and tested the localization accuracy with stationary and mobile blind nodes. The experiment consisted of ten trials of 100 s each. The mobile blind node moved based on the RWP model with random speeds between of 1 to 2 m/s and zero pause time and with $w_r = 20$ and $w_l = 20$, whereas the stationary blind nodes stayed at the center of the testing area with $w_r = 50$ and $w_l = 50$.

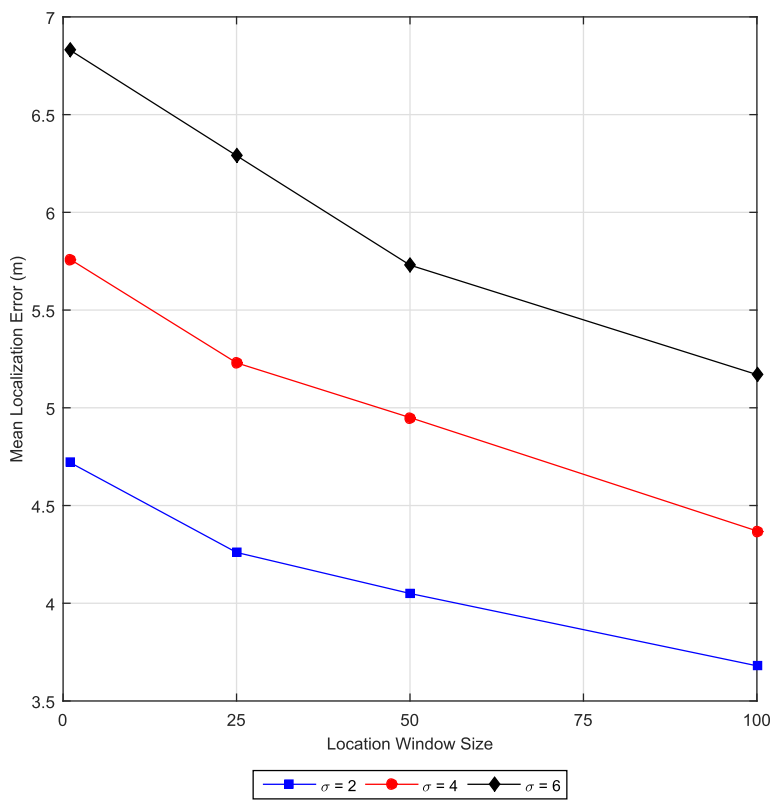

**Fig. 6** Effect of estimated location window size on the mean localization error when the reference nodes are mobile and the bind node is stationary

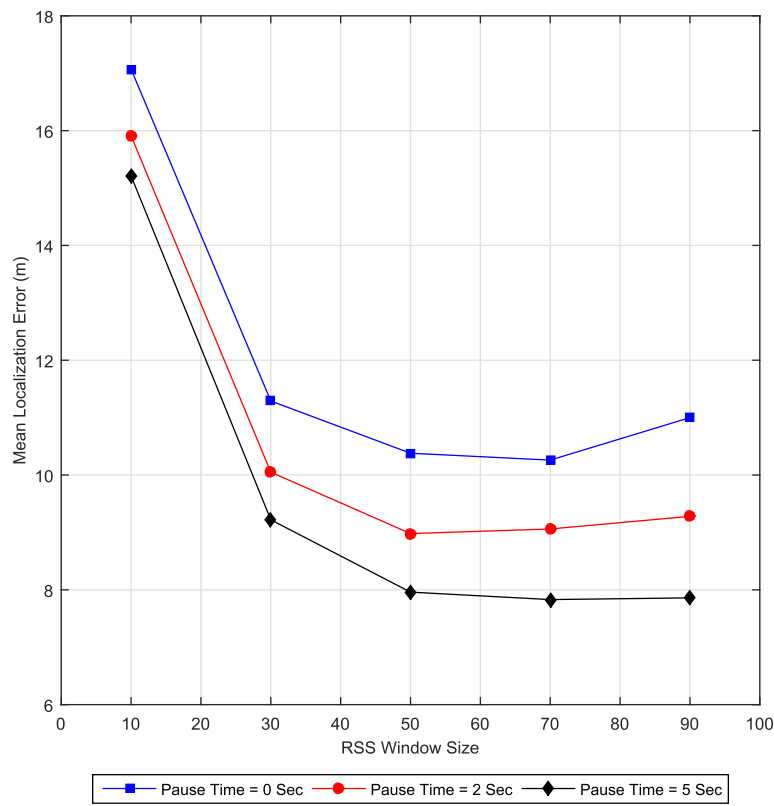

**Fig. 7** Effect of RSS window size and mobility pause time on the mean localization error when all nodes are mobile

Figure 8 shows the obtained results for the stationary and mobile blind nodes at different grid sizes. The results suggest that reasonable cost-effective localization accuracies can potentially be achieved using this strategy.

## 4 Empirical testing results

To validate the simulation results obtained in Section 3 and to verify the practical feasibility of DALIS for indoor location identification with stationary reference nodes, a small-scale experimental testing was conducted in a real indoor-obstructed environment via two sets of experiments. In the first experiment, a stationary blind node testing included a standstill and an in-place random moving and rotating. In the second experiment, a mobile blind node at a walking-speed testing was conducted along a predefined path within the testing environment. This corresponds to Section 3.3 of the simulation study.

### 4.1 Experimental testing setup

Due to the scarcity of existing support for ad hoc mode in the commercial off-the-shelve Wi-Fi devices, the ESP-8266 open-source Wi-Fi module [41] was used and a special firmware was developed to use it as a DALIS-based reference node. In addition, a simple application was developed for an Android mobile phone to use it as a DALIS-based blind node. For the purposes of this experimental testing, an HTC M8 mobile phone was used as a blind node. Figure 9 shows snapshots of the hardware and software tools developed for the experimental testing. The ESP8266 module with a developed power adapter circuit (left) allows for a fast reference node deployment. The mobile application (right) registers and logs the RSS levels as they are reported by the device radio and the estimated PLE sent by the reference nodes. Then, based on the collected data, the distance to each reference node and the corresponding relative location estimation are calculated.

Since the ESP8266 module does not support ad hoc mode, it was configured to function as an access point. However, the standard access point operation does not allow any data communication with other access points over the wireless link or with client devices (e.g., the blind node) until they associate with the access point.

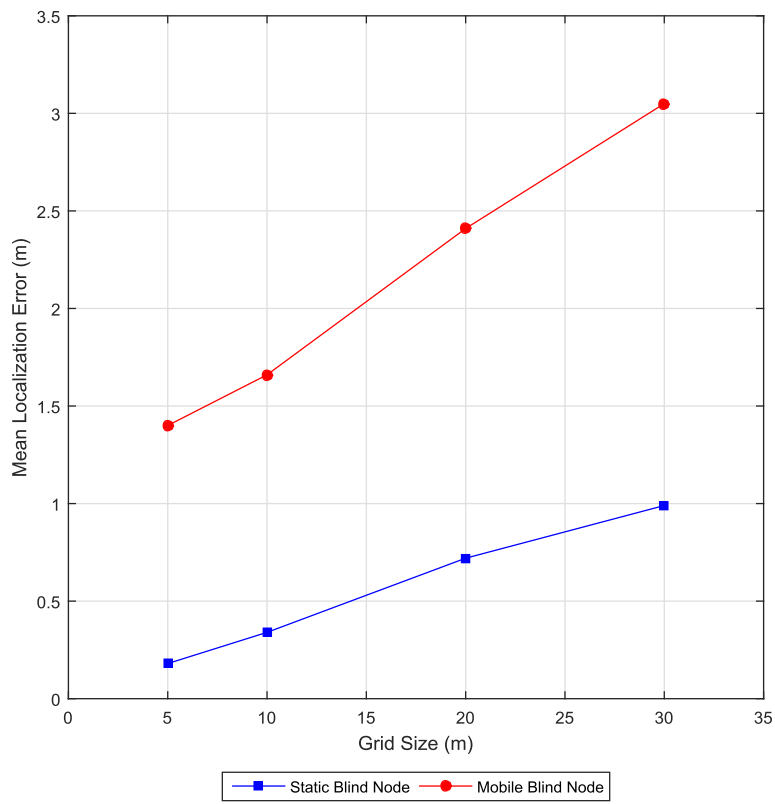

**Fig. 8** Effect of grid size of the distribution of the reference nodes on the mean localization error

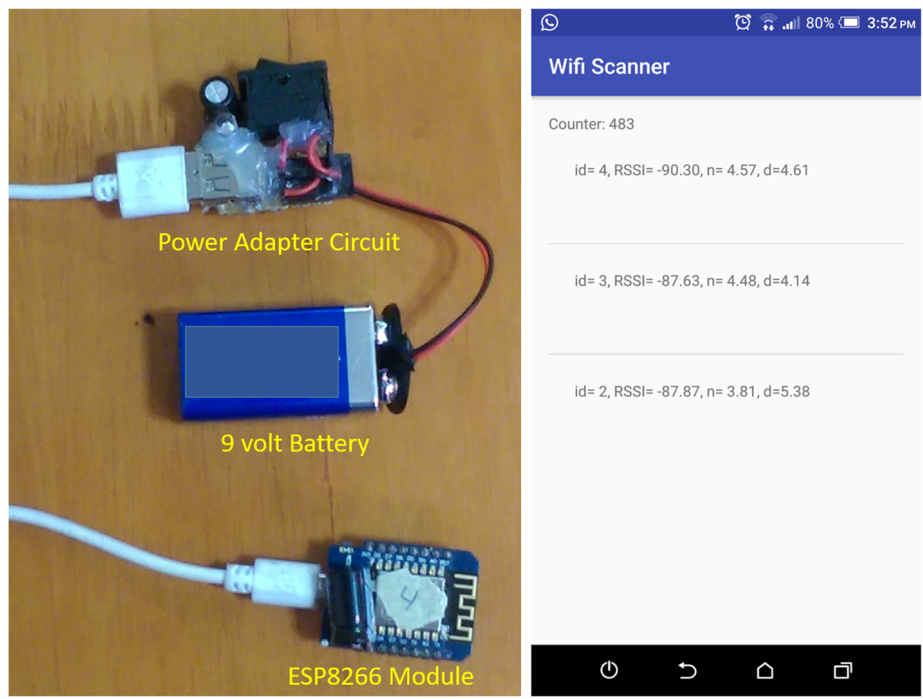

**Fig. 9** Hardware (left) and software (right) tools developed for experimental testing

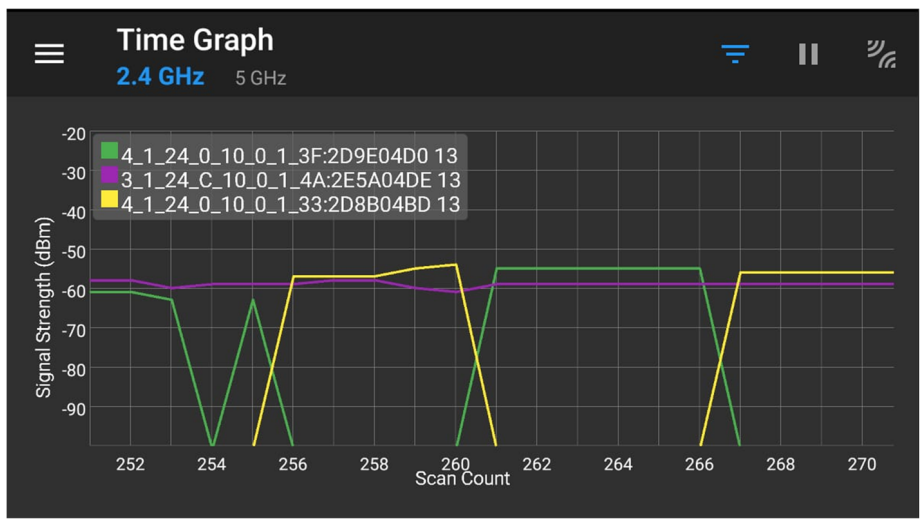

**Fig. 10** A snapshot of encoded SSIDs of two reference nodes

Furthermore, the standard Wi-Fi client device is not allowed to associate with more than one access point at a time. On the other hand, in DALIS, the blind node is required to receive data from all nearby reference nodes without having to associate with any. Therefore, a special firmware was developed for the ESP8266 modules in order to allow it to exchange DALIS parameters via especially encoded Service Set ID (SSID) within the standard beacon frame. Each SSID consists of a node ID, the coordinates, and the latest estimation of the PLE. Thus, the SSID of each reference node may change at any moment as one or more parameters it contains are updated. The only parameter that may not change is the node ID. The blind node application was also made to decode such SSIDs and distinguish them via their node IDs. Figure 10 shows a snapshot of encoded SSIDs for two reference nodes over a time graph captured using the "WiFi Analyzer" open-source mobile application [42]. Note that the leftmost digit of the SSID represents the node ID. Thus, reference node with ID 4 used two different SSIDs over the captured period, which indicates that at least one parameter (e.g., estimated PLE) has been modified, whereas reference node with ID 3 used one SSID. The reference nodes were configured to calculate the PLE based on an RSS averaging window size of 50 (i.e., $w_r = 50$).

The testing environment used was a technical workshop, which represents an obstructed open space indoor environment. A snapshot of the testing area and setup is shown in Fig. 11. Three reference nodes were deployed at three corners of a 4 m × 4 m square at a height of about 1.75 m, representing a grid-based setup as explained in Section 3.3. The three reference nodes were located at ($x$, $y$) coordinates (4, 0), (4, 4), and (0, 4), respectively. Figure 11 also shows the spots used for stationary testing as well as the path used for mobility testing. The stationary testing spots are located at (2, 0), (2, 1), (2, 2), (2, 3), and (2, 4), and the mobility testing path forms a rectangle around the middle set of benches with two opposite corners located at (− 0.5, 2) and (4.5, 4.5).

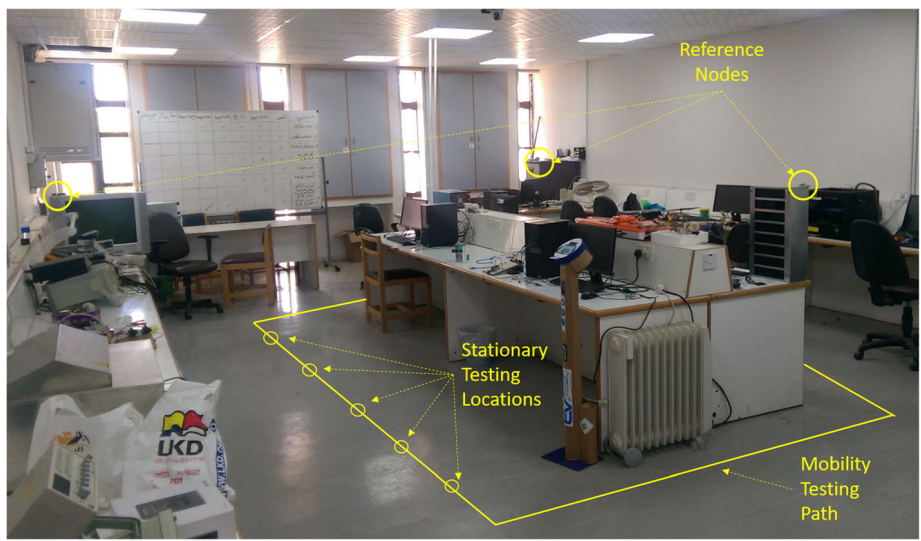

**Fig. 11** Indoor environment used for experimental testing

### 4.2 Case 1: Stationary blind node

In this experiment, a stationary blind node location identification testing was conducted using two scenarios: standstill and in-place rotating.

#### 4.2.1 Standstill blind node experiment

In this scenario, a blind node was held without any movement for 2 to 3 min in each of the five abovementioned stationary testing spots and the location was estimated after each new RSS sample is passed from the device radio. These location estimates were enumerated and statistically analyzed to assess the average and distribution of the localization accuracy. Figure 12 shows the overall error probability distribution at the testing locations altogether with an MLE of 1.6 m and 90% of the errors are below 2 m.

To have a closer look at the details, let us examine the testing spot located at (2, 2) and see how the system actually behaved over time. Figure 13 shows the instantaneous estimated PLE, $\hat{n}$, estimated by each reference node along with the localization error experienced by the blind node over 200 samples. The mobile application received approximately

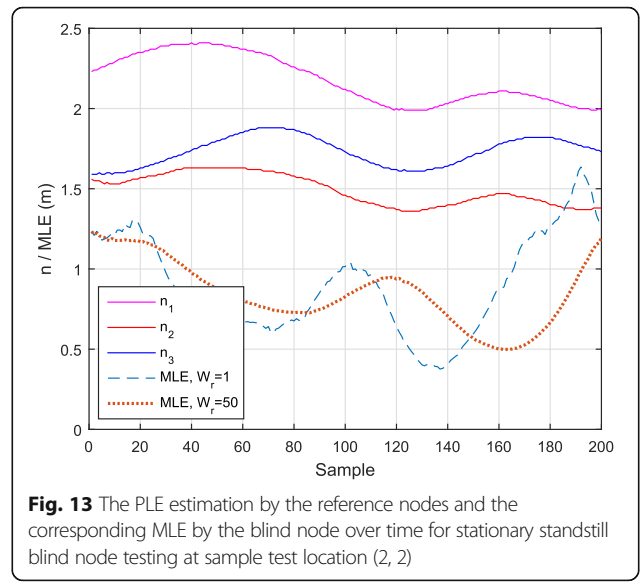

Fig. 13 The PLE estimation by the reference nodes and the corresponding MLE by the blind node over time for stationary standstill blind node testing at sample test location (2, 2)

one sample RSS from the device radio every 1 to 2 s. Figure 13 demonstrates how the PLE at each reference node adapts smoothly and independently with the dynamics of the environment. This, along with the RSS variations at the blind node's device radio (Fig. 14), causes some fluctuation in the estimated location, represented by the instantaneous localization error. Note how the RSS averaging in Fig. 14 filters out the sudden instantaneous variations in the environment and variations in the location estimation (Fig. 13).

The spatial distributions of the instantaneously estimated locations, using different RSS averaging window sizes, are shown in Fig. 15. As the window size increases, the estimated locations become less scattered around the mean estimated location, calculated as the centroid of all estimated locations and represented as a triangle in Fig. 15.

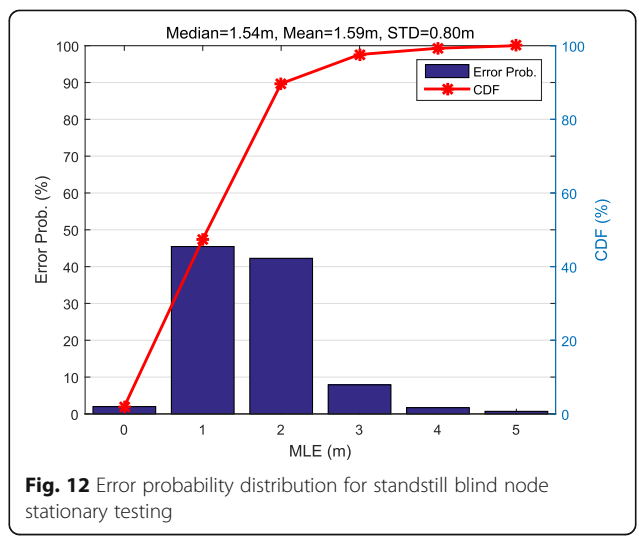

Fig. 12 Error probability distribution for standstill blind node stationary testing

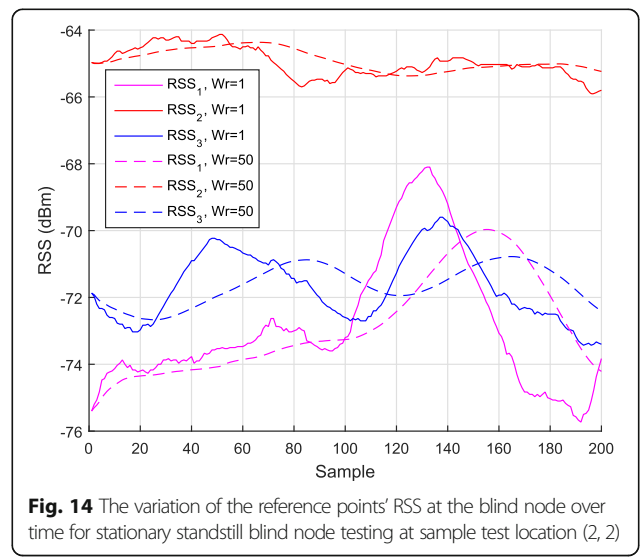

Fig. 14 The variation of the reference points' RSS at the blind node over time for stationary standstill blind node testing at sample test location (2, 2)

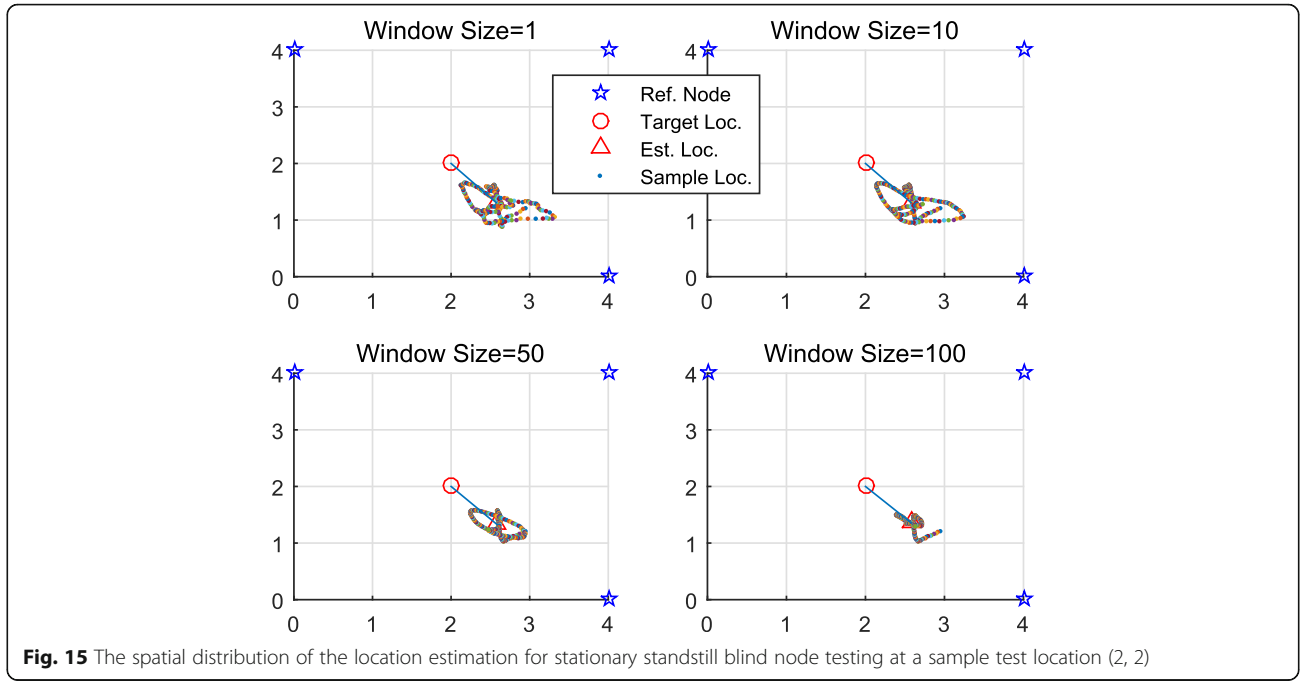

**Fig. 15** The spatial distribution of the location estimation for stationary standstill blind node testing at a sample test location (2, 2)

#### 4.2.2 Moving in-place blind node experiment

In this experiment, the scenario in Section 3.2.1 was repeated except that the blind node was slightly moving up and down and slowly turning left and right and rotating around itself in a random fashion. This resembles a more practical scenario of a person holding a mobile phone and naturally looking around for a direction to move in than the standstill scenario.

Figures 16 shows that the MLE of the moving in-place scenario is about 15% larger than that of the standstill scenario and its error distribution is somewhat worse as only about 75% of the errors are less than 2 m (versus almost 90% for standstill). However, the moving in-place scenario obtained 9% errors with less than 1 m (versus only 2% for standstill). Overall, almost all errors (~ 97%) in both scenarios are less than 3 m. The moving in-place seems to benefit from the rotation by neutralizing the effect of the user body that may be blocking the signal of certain reference nodes. This gives the blind node a better chance of getting a line-of-site signal from all reference nodes.

To examine the results obtained at sample testing point located at (2, 2) again, Fig. 17 depicts that even though the environment is moderately stable, represented by the estimated PLE's by the reference nodes, the instantaneous localization error is not, and it goes

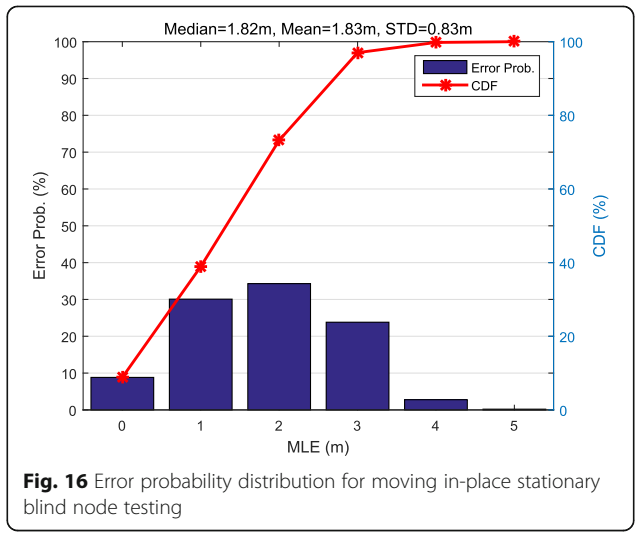

**Fig. 16** Error probability distribution for moving in-place stationary blind node testing

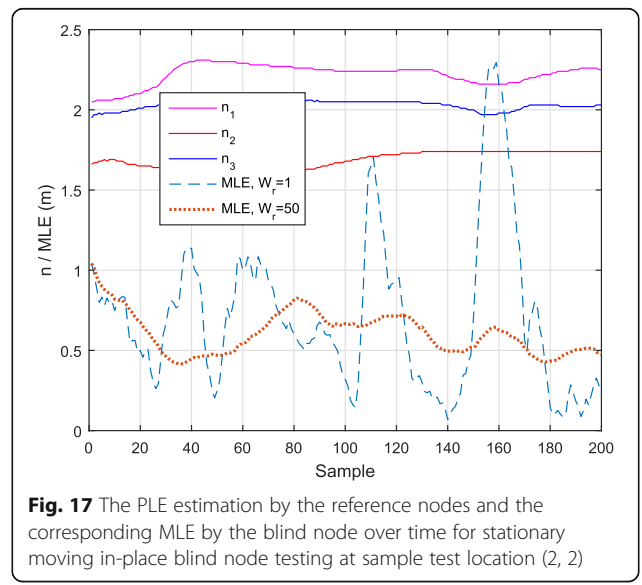

**Fig. 17** The PLE estimation by the reference nodes and the corresponding MLE by the blind node over time for stationary moving in-place blind node testing at sample test location (2, 2)

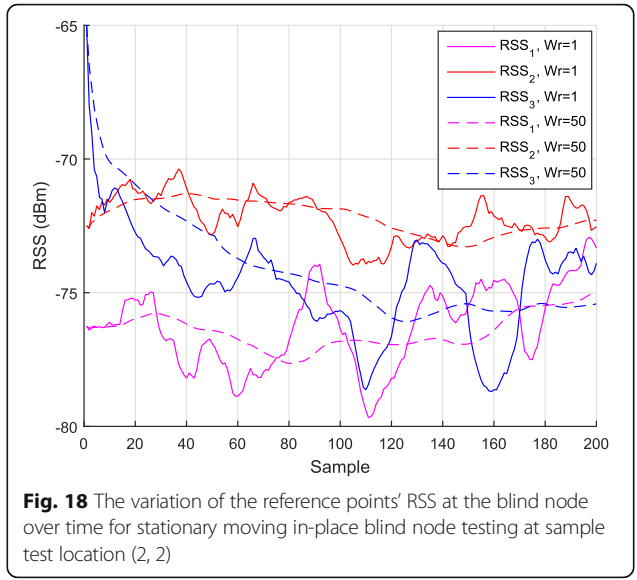

Fig. 18 The variation of the reference points' RSS at the blind node over time for stationary moving in-place blind node testing at sample test location (2, 2)

through quasi-periodic slight ups and downs. This is caused by the continuous in-place movements of the blind node. This behavior can also be noticed on the reference nodes' RSS values at the blind node, as shown in Fig. 18. However, the RSS averaging at the blind node can still smooth out these variations and provide more stable location estimation.

Figure 19 shows that, even though the instantaneous location estimations are more scattered than those of the standstill scenario, the RSS averaging by the blind node can still filter out the scattering and provide more consistent location estimation.

### 4.3 Case 2: Mobile blind node

In this experiment, a simple blind node tracking test was conducted by moving the blind node along a predetermined path (as explained in Section 3.1) and comparing it with the path estimated by the blind node.

The result of the experiment is shown in Fig. 20. Even though the estimated tracking path does not look very much like the traversed path, it still covers the most part of it within 2 m accuracy, especially when an RSS window size of 10 is used. Since there is no other precise tracking system that can be used as a reference, there is no one-to-one correspondence between the actual locations and estimated locations. Therefore, it is infeasible to perform any statistical analysis or comparison and hence visual inspection can be only used.

As explained in Section 2, with node mobility, using relatively large window sizes can degrade the localization performance since early RSS samples expire and become no longer valid. This is clearly depicted in Fig. 20. $w_r = 5$ and $w_r = 10$ provide an incremental improvement to the results, whereas $w_r = 25$ degrades it significantly.

### 4.4 Limitations of the current experimental testing

The system developed for experimental testing is fairly simple and was mainly designed to prove the basic concepts of the proposed methodology. There are, in fact, a number of limitations the system suffers from. One limitation is that the radiation pattern of the ESP8266 surface mount antenna is not uniform. Figure 21 shows the radiation pattern of the ESP8266 antenna that was experimentally constructed, as reported in [43]. There is approximately 6 dB difference between the minimum and maximum signal levels at angles 70° and 250°, respectively. This would definitely confuse the PLE calculations at the reference nodes and would add considerable noise to the distance estimation at the blind node, depending on the angle of incidence. To solve this problem, an external antenna with a sufficiently uniform radiation pattern has to be used, which mandates some hardware modifications to the module. Another software-related limitation is the inability of a simple mobile application to access the low-level Wi-Fi radio parameters. This is needed to control the RSS averaging, which was shown to have a significant impact on the localization performance. Existing operating systems usually perform some low-level RSS averaging before the scanning results are reported to the user-space. The typical beacon rate of a Wi-Fi access point is 10 frames per second, whereas existing commercial mobile phones take a few seconds before updating the scanning results to the user space, including the RSS signal strength [44]. Hence, having a system-level access to the Wi-Fi radio can help improve the localization and tracking performance significantly. In addition, in real-life location identification and navigation application, real mapping information (e.g., wall, walking paths, and hallways) can also be used to improve the system performance.

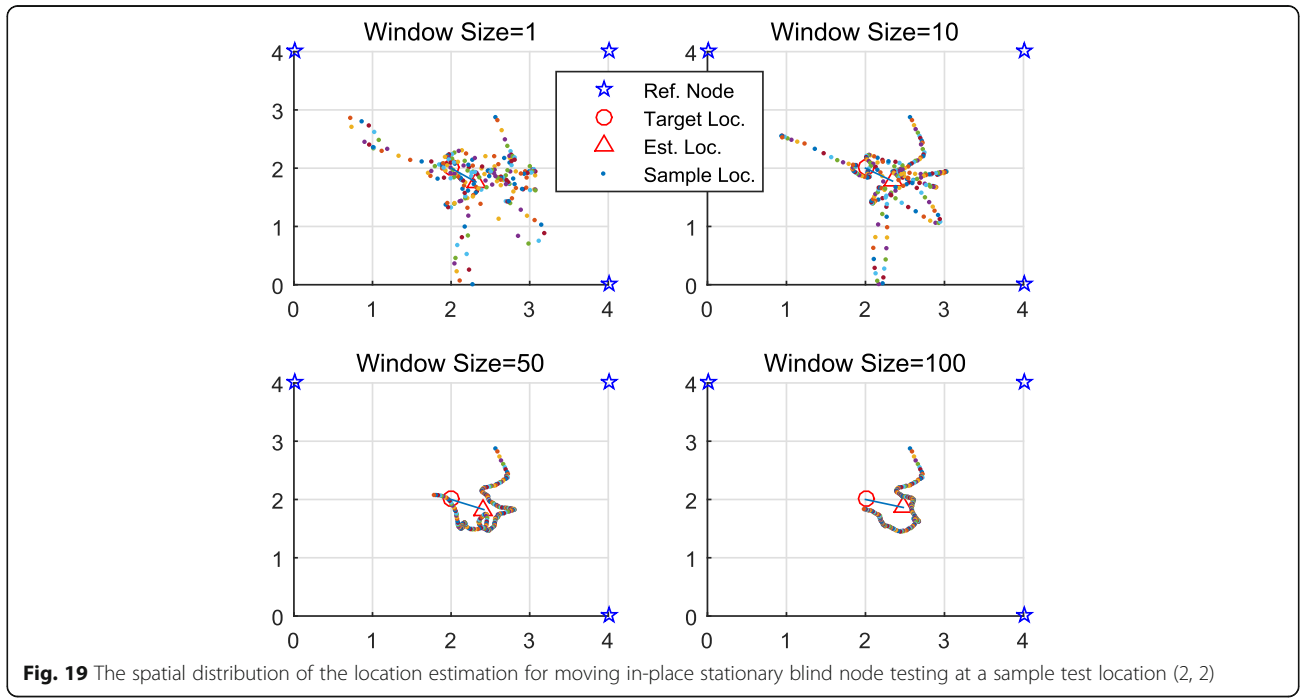

Fig. 19 The spatial distribution of the location estimation for moving in-place stationary blind node testing at a sample test location (2, 2)

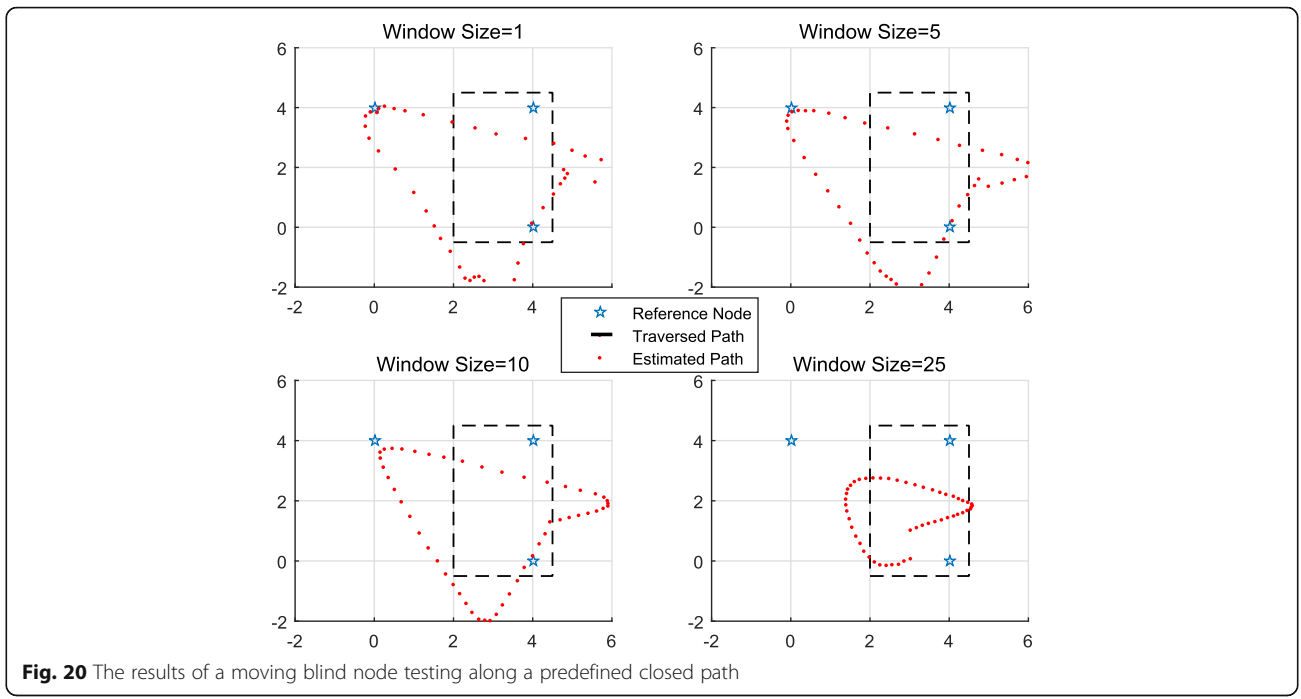

**Fig. 20** The results of a moving blind node testing along a predefined closed path

As a conclusion, in spite of the abovementioned limitations the experimental testbed suffered from, on top of using two different types of Wi-Fi devices, the results of the experimental testing are fairly acceptable, confirming the high potential the proposed location identification methodology has.

### 4.5 Comparison with state of the art

Comparing the accuracy of different indoor localization algorithms is a challenging task to do, given the diversity in hardware components and devices and in software implementations. Therefore, the comparison depends mainly on how well the authors describe the test scenarios and on how the reader interprets that. Table 5 compares the empirical results of the proposed DALIS algorithm to the state-of-the-art methods discussed earlier within the related work. In summary, as far as the localization accuracy is concerned, the performance of DALIS is better than most of the state-of-the-art techniques. Furthermore, DALIS is unique in being the only fully distributed and simple pure model-based technique for navigation and location identification. The only method that achieves higher accuracy than DALIS is IMM-EKF, which is a very complicated method that uses a centralized server to process and analyze the RSS data and track the mobile target. Thus, the advantages of DALIS are its simplicity in terms of the required calculations and in involving each device in performing its own tasks. On top of that, privacy of the mobile device is maintained.

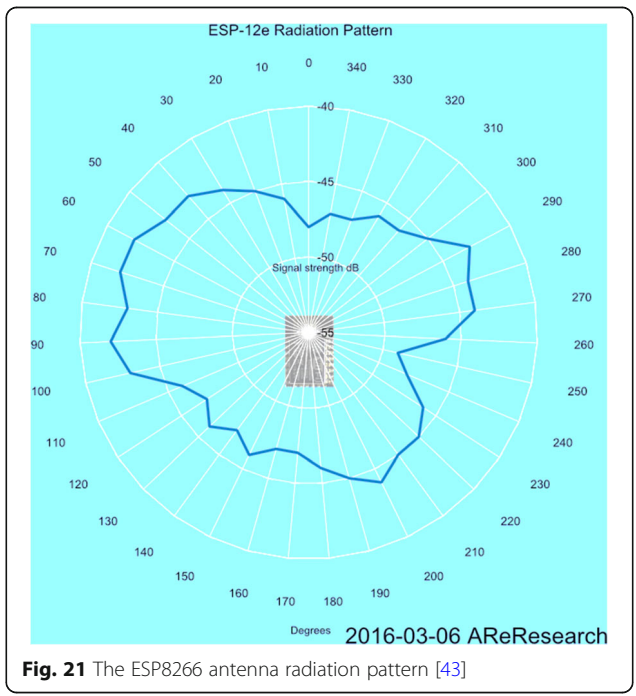

**Fig. 21** The ESP8266 antenna radiation pattern [43]

Table 5 A comparison between the proposed method and the recent state-of-the-art model-based methods

| Method | Model | Architecture | Scenario | MLE | < 3 m |
|---|---|---|---|---|---|
| A self-adaptive model-based [33] | Modified dual-model | Centralized client-server (localization and tracking) | Single room, stationary | 2–3 m | 70% |
|  |  |  | Multi-room, stationary | 3–4 m | 70% |
| MFAM [34] | Modified dual-mode model with multi-frequency | Centralized client-server (localization and tracking) | Multi-room, stationary | 2.16 m | NA |
| IMM-EKF [35] | Dual-slope, interacting multi-model with Kalman filters | Centralized client-server (localization and tracking) | Multi-room, mobile | 0.19 m | 100% |
| DALIS (proposed) | Pure model-based | Distributed ad hoc (pseudo-GPS location identification) | Single room, stationary, standstill | 1.59 m | 98% |
|  |  |  | Single room, stationary, moving in-place | 1.83 m | 97% |

## 5 Discussion

In light of the numerous technical challenges and limitations that face traditional infrastructure-based indoor localization systems such as costly deployment and centralization; poor reliability, accuracy, and adaptability to the RF environment changes; delayed; complexity; labor-intense operation; and security concerns, this paper presents an innovative core solution to most of such challenges and deficiencies.

The key principle of the proposed methodology is to keep the system as simple and robust as possible. The distributed and ad hoc nature of the mechanism makes easier and faster to deploy and more reliable than infrastructure-based counterparts. In addition, its simplicity makes easy to install and operate in resource-limited devices because the algorithm is based on closed-form mathematical expressions, in addition to limited storage and averaging operations of the RSS samples or estimated locations. Thus, its overall computational complexity is comparable to that of the real-time GPS-based navigation systems. Furthermore, it collaborative nature allows it to utilize whatever devices are available to achieve the best possible performance.

The simulation and experimental testing results, reported earlier in the paper, substantiate that the proposed methodology can potentially provide practical location identification for blind devices in a number of scenarios and for several critical applications. According to the results, the most practically durable scenario is when the reference nodes are stationary and the blind node is moving at the normal human walking speed, which makes it suitable for large-scale indoor navigation and location-based services such as in shopping malls, hospitals, museums, and city halls. However, if the reference nodes are mobile, such as in collaborative ad hoc mobile phone localization, the possible continuous mobility of all nodes makes the system less durable for the general navigation applications. On the other hand, it can possibly be acceptable for certain critical applications such as search-and-rescue operations, where rough location estimation can suffice with the use of other sensing capabilities such as vision and hearing that can help pinpoint the target location.

To illustrate the potential of the proposed approach, assume that a swarm of small and low-cost unmanned aerial vehicles (UAV), such as quadcopter drones, are equipped with GPS-enabled autopilot systems and Wi-Fi radios. These drones can now be easily programmed to autonomously position themselves at a proper GPS-clear altitude and within radio communication range of one another. This swarm of drones can serve, based on the proposed mechanism, as a pseudo-GPS system for ground-based blind mobile devices or stations within a GPS-obstructed indoor or outdoor environment. Several applications can benefit from this setup such as a search-and-rescue operation (see [45] as an example) by either human rescuers or unmanned ground vehicles (UGV) such as mobile robots (see [46] as an example). Similarly, low-cost Wi-Fi modules can be deployed in an indoor environment at known locations and with a proper density to facilitate an indoor pseudo-GPS system for indoor navigation and location-based services. Perhaps, the most challenging scenario for the proposed mechanism is to use location-aware mobile devices as reference nodes, which makes it less accurate and less robust. However, in situations where no better other choices exist, a rough location estimation is better than none. In addition, intelligent software and mapping techniques can be employed in such case to enhance the efficiency at higher computational costs.

## 6 Conclusions

The current research work proposes a non-traditional novel GPS-like location identification strategy for indoor or GPS-obstructed environments. The proposed strategy is simple, distributed, and collaborative, and is of low cost, yet efficient and robust with the potential to become the best alternative to most existing costly infrastructure-based strategies. The proposed approach has low complexity, and it can adapt to dynamically changing RF environments to provide real-time location information. The simulation and empirical testing revealed that the proposed strategy suits a large number of potential scenarios and critical applications and has the potential to provide location information with an accuracy of less than 3 m.

The main contribution of this work is a low-cost distributed and adaptive indoor location identification and navigation system for mobile devices. This system can be easily deployed using low-cost IoT Wi-Fi modules that can be used as beacon nodes in a pseudo-GPS system for location referencing. The mobile blind node uses the beacon packets to independently identify its own location and navigate through the environment in real time.

As a future work to this research, some limitations, such as device heterogeneity and its impact on the location identification accuracy, are to be investigated. In addition, a large-scale implementation and integration with the real mapping information is to be considered.

#### Acknowledgements
The authors would like to thank the Deanship of Scientific Research at the Jordan University of Science and Technology for funding this research project.

#### Funding
This research was funded by the Deanship of Scientific Research, Jordan University of Science and Technology, Research Grant Number 20130172.

#### Authors' contributions
The authors declare that they all contributed to the manuscript. All authors read and approved the final manuscript.

#### Competing interests
The authors declare that they have no competing interests.

#### Publisher's Note
Springer Nature remains neutral with regard to jurisdictional claims in published maps and institutional affiliations.

#### Author details
[1]Department of Network Engineering and Security, Jordan University of Science and Technology, P. O. Box 3030, Irbid 22110, Jordan. [2]Department of Computer Science, Jordan University of Science and Technology, P. O. Box 3030, Irbid 22110, Jordan. [3]Department of Computer Engineering, Jordan University of Science and Technology, P. O. Box 3030, Irbid 22110, Jordan.